\documentclass[doublecol]{epl2} 

\usepackage{amsmath,amssymb}

\newcommand{\toyLIFE}{{\fontfamily{qcr}\selectfont t\!\raisebox{-.1em}{o}\!y}%
{\fontfamily{phv}\selectfont \-LIFE}}

\title{Statistical theory of phenotype abundance distributions: a test through exact enumeration of genotype spaces}
\shorttitle{Statistical theory of phenotype abundance distributions} 

\author{%
Juan Antonio Garc\'{\i}a-Mart\'{\i}n\inst{1,2,3} \and
Pablo Catal\'an\inst{1,4} \and
Susanna Manrubia\inst{1,2} \and
Jos\'e A. Cuesta\inst{1,4,5,6}
}
\shortauthor{J. A. Garc\'{\i}a-Mart\'{\i}n \etal}

\institute{                    
\inst{1} Grupo Interdisciplinar de Sistemas Complejos (GISC), Madrid, Spain\\
\inst{2} Programa de Biolog\'{\i}a de Sistemas, Centro Nacional de
Biotecnolog\'{\i}a (CSIC), Madrid, Spain\\
\inst{3} Bioinformatics for Genomics and Proteomics, Centro Nacional de
Biotecnolog\'{\i}a (CSIC), Madrid, Spain \\
\inst{4} Departamento de Matem\'aticas, Universidad Carlos III de Madrid, Legan\'es, Madrid, Spain\\
\inst{5} Instituto de Biocomputaci\'on y F\'\i sica de Sistemas Complejos
(BIFI), Universidad de Zaragoza, Spain\\
\inst{6} UC3M-BS Institute of Financial Big Data (IFiBiD), Universidad Carlos
III de Madrid, Getafe, Madrid, Spain
}

\abstract{The evolutionary dynamics of molecular populations are strongly
  dependent on the structure of genotype spaces. The map between genotype and
  phenotype determines how easily genotype spaces can be navigated and the
  accessibility of evolutionary innovations. In particular, the size of neutral
  networks corresponding to specific phenotypes and its statistical counterpart,
  the distribution of phenotype abundance, have been studied through multiple
  computationally tractable genotype-phenotype maps. In this work, we test a
  theory that predicts the abundance of a phenotype and the corresponding
  asymptotic distribution (given the compositional variability of its genotypes)
  through the exact enumeration of several GP maps. Our theory predicts with high
  accuracy phenotype abundance, and our results show that, in navigable genotype
  spaces ---characterised by the presence of large neutral networks---, phenotype
  abundance converges to a log-normal distribution. 
}

\begin{document}

\maketitle

\section{Introduction}

How the genetic information maps into functional phenotypes (the so-called
genotype-to-phenotype, or GP, map) critically conditions the dynamics of
evolution~\cite{cowperthwaite:2007,aguirre:2018}. Genotypes encode the information
to generate phenotypes and {in} the process of replication undergo all sorts of
mutations. The second basic mechanism of evolution, selection, acts upon
phenotypes. Standard approaches to evolutionary dynamics have traditionally
overlooked the fact that genotype and phenotype are connected through very
complex mechanisms, and that the latter may have strong effects on the dynamics.

Genotype spaces can be depicted as networks, with nodes representing genotypes
and links joining pairs of genotypes mutually accessible through a mutation.
Phenotypes are then subsets of nodes in this network, and the GP map describes
their distribution in genotype space. As selection acts on phenotypes, evolution
within a connected component of a phenotype is neutral (or nearly so). For this
reason, they are referred to in the literature as neutral networks
(NNs)~\cite{schuster:1994,bornberg-bauer:1997}. A characteristic feature of all
known GP maps is the strongly heterogeneous distribution of the abundance (number
of nodes) of their NNs~\cite{wagner:2011,ahnert:2017}. A few NNs are huge,
typically percolating the whole genotype space, whereas most of them are small.
This has evolutionary implications. First of all, the existence of huge NNs
endows populations with a high genomic variability without bearing any selective
cost. Secondly, most phenotypes are not accessible for
entropic reasons~\cite{schaper:2014,dingle:2015,catalan:2017}. Besides, large
NNs are so interwoven that virtually any pair of them are connected to each
other, thus facilitating the search for
phenotypes~\cite{gruner:1996b,fontana:1998b}. Under this paradigm, evolution is
both robust and innovative.

\begin{figure*}
\includegraphics[width=170mm,clip]{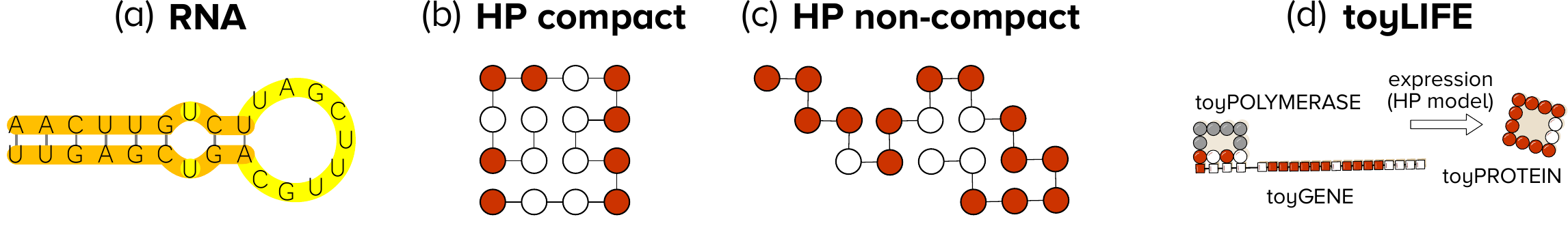}
  \caption{Schematic representation of the different GP maps exhaustively studied
    in this work. (a) In RNA, sequences are folded to minimum free energy
    secondary structures that define the phenotype; (b) in the compact version of
    the HP model, hydrophobic (H, white circles) and polar (P, red circles)
    residues adopt the minimum compact energy configuration; (c) in non-compact
    HP, sequences are assigned to self-avoiding walks of minimum energy;
    (d) \toyLIFE{} is a multilevel GP map with HP-like sequences that {code} for
    compact HP interacting proteins. {Phenotype definitions can be found in Appendix~A.}}
\label{fig:Fig1}
\end{figure*}

Given the complexity of GP maps, we need to uncover and characterise as many
general features as possible. One of them is the abundance distribution of NNs.
The first studies of this distribution often relied on random samplings of the
genotype space and considered relatively short RNA
molecules~\cite{schuster:1994b,gruner:1996}. These are chains of a two- to
four-letter alphabet (A, U, C, G or a subset of those), whose phenotype is
identified as a minimum-free-energy folding (secondary
structure)~\cite{schuster:2006}. Results pointed to a fat-tailed, decaying
distribution
\cite{gruner:1996,stich:2008,ferrada:2012,cowperthwaite:2008,aguirre:2011}
---although whether exponential, power-law, or otherwise is far from clear.
Later studies of longer molecules (up to 126 letters long) show bell-shaped
abundance distributions instead~\cite{dingle:2015}.

The first theoretical model addressing this question considered a set of binary
sequences with a specific GP mapping rule~\cite{greenbury:2015}: the abundance
distribution was an unequivocal power law. Later, it was pointed out that two
different kinds of distributions ---power-law and log-normal--- are
possible~\cite{manrubia:2017}. The argument relies on the existence of sites
showing low and high compositional variability within a phenotype. Power laws
are expected when these positions occupy fixed sites, whereas log-normals arise
if their location is {not fixed, so that counting the number of arrangements of
them in the sequence yields a combinatorial factor}. In the case of RNA sequences,
low/high variability sites are associated to paired/unpaired nucleotides in the 
folded structure. A combinatorial calculation of the distribution of paired and
unpaired sites can be carried out exactly (see~\cite{cuesta:2017} and references
therein) and shown to be normal. As the number of low variability sites can be
related to the logarithm of the phenotype abundance, the resulting distribution
turns out to be log-normal. As a matter of fact, since not only paired sites, but
any other structural feature of the folded chain can be shown to have a normal
distribution, the argument can be extended even if site variability is affected
by other structural elements. The log-normal prediction is thus expected to be
quite robust.

\section{Versatility of a site}

An alternative way to look at the problem of estimating phenotype abundance was
suggested in the discussion of~\cite{manrubia:2017}. If, for a given phenotype,
a variable $v_i$ could measure the average number of different letters of the
alphabet that show up at site $i$ of its sequences, then the abundance could be
estimated as
\begin{equation}
S_{\text{est}}=v_1v_2\cdots v_L
\label{eq:estimatedsize}
\end{equation}
if the genotype is a chain of length $L$. This definition is easy to understand
if sites are either completely neutral (any mutation maintains the phenotype,
$v_i=k$, {the size of the alphabet}) or fully constrained (any mutation changes
phenotype, $v_i=1$). In a more general case, $v_i$ would take intermediate
values.

Given that phenotypes differ in the distributions of their structural motifs,
and that the variability of a site is strongly correlated to the motif it sits
in, variables $v_i$ can be regarded as phenotype-dependent random variables that
take values from a certain distribution. Thus, by the central limit theorem
$\ln S$ will be a phenotype-dependent, normally-distributed random variable.

Here is a way to estimate one such variable $v_i$ (henceforth referred to as
\emph{versatility}) {for} an alphabet of $k$ letters. We choose a phenotype and
count in how many of its genotypes letter $\alpha$ shows up at site $i$. Let
$m_{\alpha,i}$ be that number. Then we define the versatility at site $i$
through
\begin{equation}
v_i=\frac{1}{M_i}\sum_{\alpha=1}^km_{\alpha,i}, \quad
M_i\equiv\max\{m_{1,i},\dots,m_{k,i}\}.
\label{eq:versatility}
\end{equation}
The rationale behind this definition relies on assuming that the relative
frequencies of each letter of the alphabet at each position $i$ are proportional
to the fraction of the space of genotypes associated to the phenotype. It
implicitly assumes that the most frequent letter at each position is always
characteristic of the phenotype, while other letters, appearing less frequently,
may yield sequences corresponding to different phenotypes. For example, if G
appears $m_{G,i}$ times and C appears $m_{G,i}/2$ times, other letters being
absent, the versatility of that site would be $v_i=3/2$, meaning that a half of
the mutations from G to C at that site change phenotype. When only one letter
appears, $v_i=1$, while $v_i=k$ if all letters are equally likely,
recovering the limits of simple models~\cite{greenbury:2015,manrubia:2017}.

\section{Testing the definition of versatility}

In order to show that the versatility introduced in Eq.~\eqref{eq:versatility}
is a meaningful concept, we have tested it for different GP maps (sketched in
Fig.~\ref{fig:Fig1}) regarding how well it predicts the abundance of a specific
phenotype component and its relationship with the distribution of phenotype
abundances.

\begin{figure}
\includegraphics[width=88mm,clip]{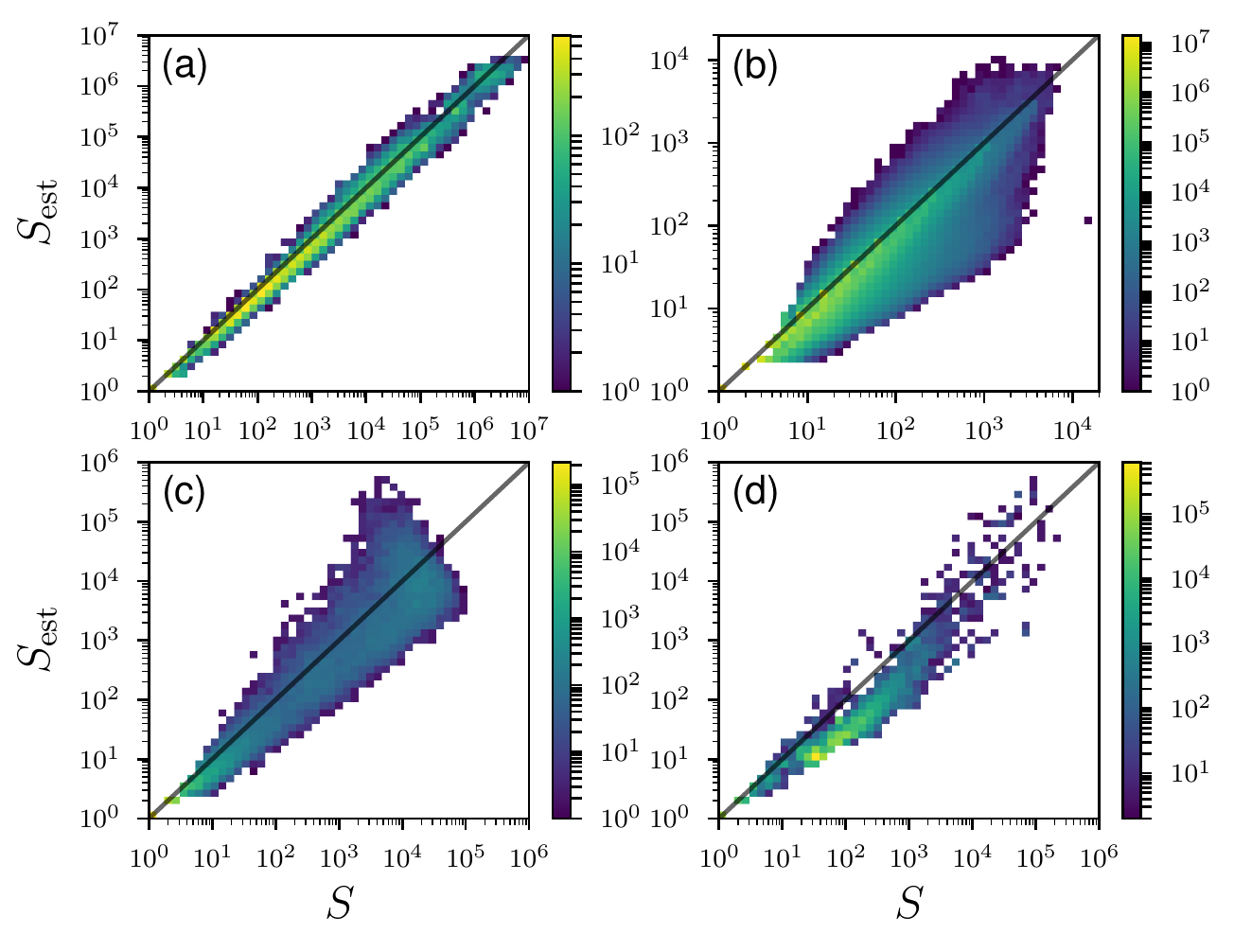}
\caption{Log-log-log histograms of the estimated abundance ($S_{\text{est}}$
  calculated as in \eqref{eq:estimatedsize}), {\it versus} actual abundance ($S$)
  of the connected components of different GP maps: (a) four-letter RNA of length
  $L=16$, (b) two-letter GC-RNA of length $L=30$, (c) compact HP model
  5$\times$6 with $U(HH)=-1$, and (d) \toyLIFE{} for two genes.}
\label{fig:Fig2}
\end{figure}

First, we have folded all RNA sequences of length $L=16$, using the Vienna
package~\cite{lorenz:2011}, and classified them according to their secondary
structures. For such a small length, phenotypes are normally fragmented into
several connected, neutral components (NCs) {of comparable size}, but exhaustively folding longer
sequences quickly becomes computationally unfeasible. Since NCs behave, to all
purposes, as independent NNs, we treat them as independent phenotypes,
regardless of whether or not they fold into the same secondary structure. Then,
we count how many sequences each NC contains (its abundance, $S$) and calculate
its site versatilities $v_i$ according to the definition~\eqref{eq:versatility}.
The product of them all yields the estimated abundance~\eqref{eq:estimatedsize}.
Fig.~\ref{fig:Fig2}(a) shows a histogram comparing actual and estimated
abundances for all the NCs, showing a remarkable agreement.
{The distinction between NCs and phenotypes becomes less relevant as the
length of genotypes grows, as discussed later (see also Appendix~B).}

A variant of this model is made of RNA sequences containing only two
complementary bases, for example G and C (GC-RNA). A two letter alphabet allows
us to study sequences almost twice as long with a similar computational
effort~\cite{gruner:1996b}. We have repeated the previous analysis for GC-RNA
sequences of length $L=30$, and plotted the result in Fig.~\ref{fig:Fig2}(b).
Fragmentation is more frequent in this model, and NCs are generally smaller. This
is why their number is so high and why they are so dispersed in
Fig.~\ref{fig:Fig2}(b). Also, the largest NCs are three orders of magnitude smaller
than those of four-letter RNA sequences. For this model, the versatility of paired
sites is strictly 1 because any mutation in such a pair will break the link.
Unpaired sites do not have much more freedom either, because a mutation can often
create a new link and change the folding. In spite of these constraints,
Fig.~\ref{fig:Fig2}(b) shows a clear correlation between $S$ and $S_{\text{est}}$,
with the overwhelming majority of NCs near the diagonal.

The third GP map that we have analysed is the HP model for lattice
proteins~\cite{li:1996}, where a protein is represented by a self-avoiding chain
of hydrophobic (H) or polar (P) beads on a lattice. The energy of a given
configuration is calculated from a contact potential,

\begin{equation}
  \label{eq:EHP}
  E = \sum_{i<j} U(\sigma_i,\sigma_j) C_{ij}
\end{equation}
where $\sigma_i \in \{H, P\}$, $C_{ij}=1$ when $i$ and $j$ are neighbours on the
lattice (with $|i-j| \ne 1$) and $C_{ij}=0$ otherwise, and $U(\sigma_i,\sigma_j)$
specifies the interaction strength. Several different specific realisations of
the model can be found in the literature (see below). For two-dimensional square
lattices, compact and non-compact versions of the model have been studied. In
compact HP, sequences of length $L=l_1 \times l_2$ are forced to fold into
rectangular structures, while non-compact HP considers all self-avoiding walks in
the lattice. In Fig.~\ref{fig:Fig2}(c) we show the case example of compact HP
5$\times$6 with a single nonzero energy parameter, $U(H,H)=-1$ {where
phenotype is defined as the non-degenerated, minimum energy conformation (see
Appendix~A).}

Finally, we have also analysed \toyLIFE, a multilevel model of a simplified
cellular biology \cite{arias:2014,catalan:2018} in which binary sequences are
first mapped to HP-like proteins that interact between themselves, with the
genome, and with metabolites. The phenotype is defined by the set of metabolites
that a given sequence is able to catabolise. Consequently, \toyLIFE{} has a lower
genotype level, which translates into proteins (second level), whose interactions
add a third, regulatory level. This regulation is altered by the presence of
metabolites, which can be catabolised as a result, giving rise to the phenotypic
expression at this highest level. Even though the connection between genotype
sites and structural elements in this model is far from clear, versatilities can
be computed nonetheless. The estimations of phenotype abundances arising from
their values, for the case of two genes (length $L=40$), are compared with actual
abundances in Fig.~\ref{fig:Fig2}(d). We can observe a slight but systematic
underestimation of abundances. In spite of that, the correlation between $S$ and
$S_{\text{est}}$ is strong, and the cloud of points runs parallel to the diagonal.
The slight underestimation of versatility, however, does not affect the argument
leading to the log-normal abundance distribution ---only the mean and the variance
will be affected.

The prediction of phenotype abundance has been a matter of study, among others
due to its relevance for protein designability~\cite{larson:2002}, for molecular
robustness and evolvability~\cite{jorg:2008}, or in the neutralist-selectionist
controversy~\cite{dingle:2015}.
Attempts at estimating phenotype abundance have been made using compositional
entropy~\cite{li:1996,larson:2002}. However, the comparison with the predictions
obtained through site versatility reveals that versatility has a superior
performance (see Appendix~C and Fig.~\ref{fig:FigS1}).

\section{Distribution of abundance of RNA NCs}

\begin{figure}
\includegraphics[width=88mm,clip]{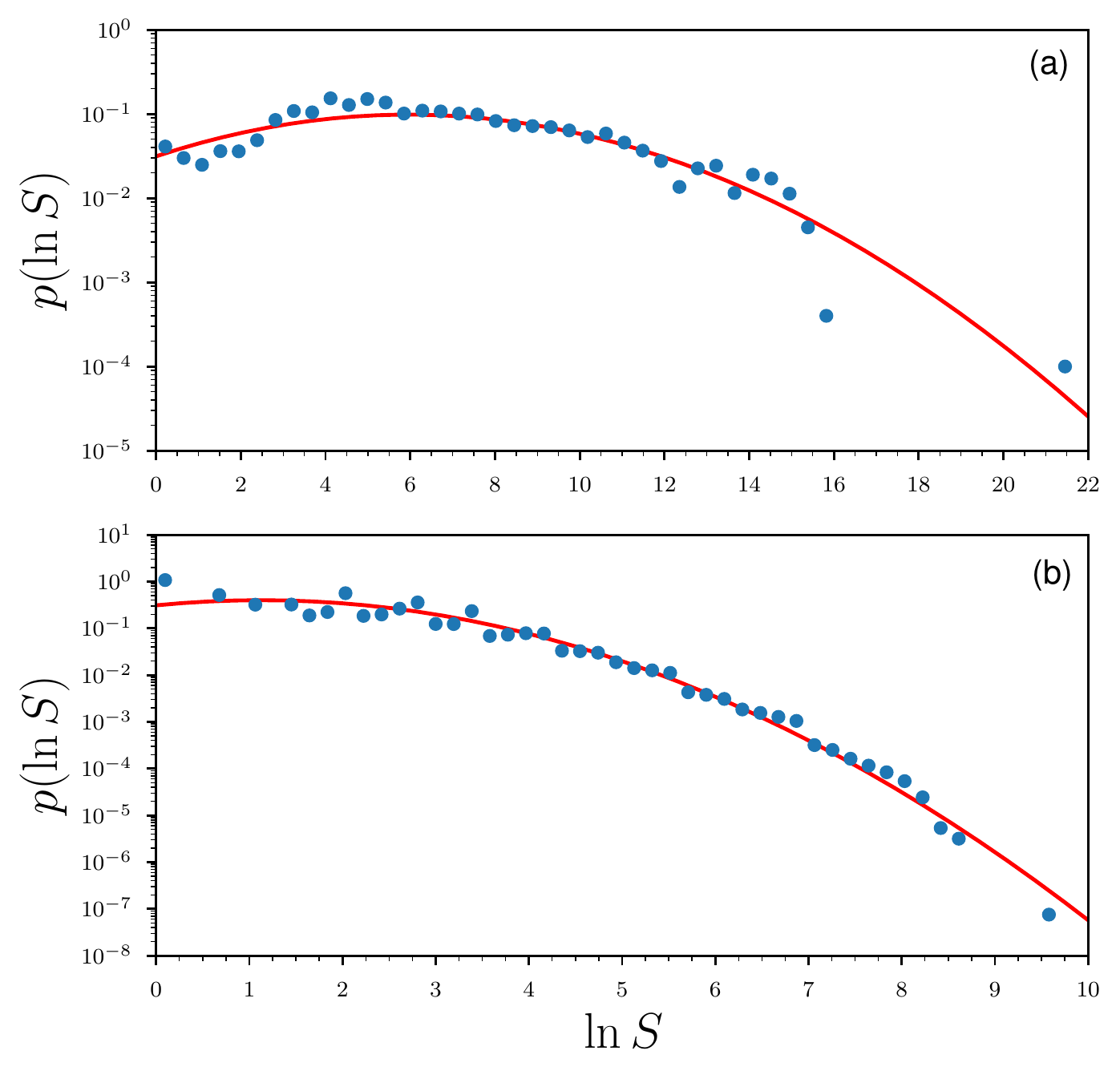}
\caption{Log-abundance distributions $p(\ln S)$ for the the NCs of (a) four-letter
  RNA sequences of length $L=16$ and (b) GC-RNA sequences of length $L=30$. Dots
  are the actual values; lines are Gaussian fits to the data.}
\label{fig:Fig3}
\end{figure}

Figure~\ref{fig:Fig3}(a) shows the distribution $p(\ln S)$ of the abundance of
RNA sequences of length $L=16$ in NCs, along with a least-squares fit of
the function $\exp[a(\ln S)^2+b\ln S+c]$, the expected asymptotic distribution
according to Eq.~\eqref{eq:estimatedsize}. The length of the sequences is too
short to exhibit a perfect Gaussian shape yet: the curve is truncated on the
left-hand side and there are deviations for small and large NCs abundances.

Though the abundance distribution of NCs for GC-RNA sequences is a decreasing
function with a fat tail (Fig.~\ref{fig:Fig3}(b)), the right tail of a log-normal
provides a good fit that captures the slight concavity of the curve. Regardless
of the alphabet size, the log-normal distribution is theoretically supported by
Eq.~\eqref{eq:estimatedsize}.

The theory developed up to now strictly applies to NCs of phenotypes. However, it
was originally inspired by studies reporting a log-normal distribution of {\it
  phenotype} abundances~\cite{dingle:2015}. Also, data corresponding to GC-RNA
phenotypes compatible with a power-law distribution~\cite{ferrada:2012} can be
fit at least equally well by a truncated log-normal such as that in
Fig.~\ref{fig:Fig3}(b). In the next section we will introduce an effective model
that will provide some insights into the specific shapes of these distributions
and clarify how the theory asymptotically applies to phenotypes.

\section{Effective two-versatility model for RNA}

\begin{figure}
\includegraphics[width=88mm,clip]{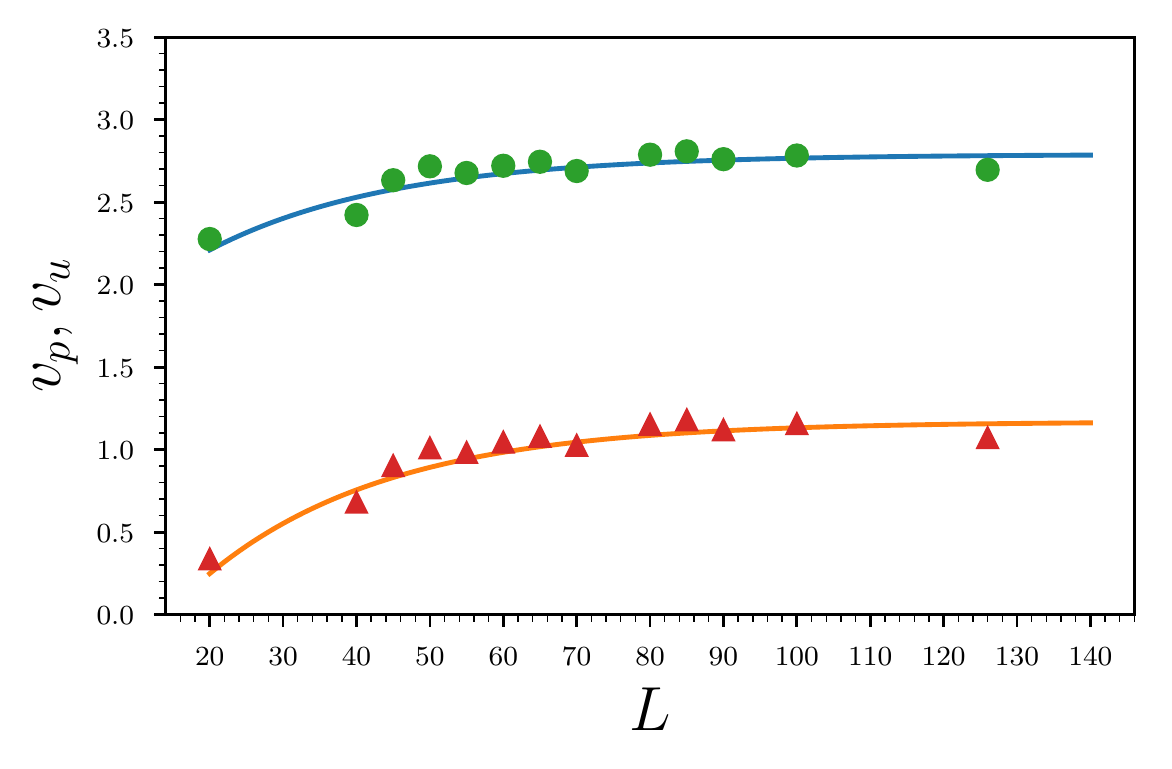}
\caption{Average versatilities of unpaired ($v_u$) and paired ($v_p$) sites
obtained by fitting a two-versatilities model to the sampled abundance
distributions of Ref.~\cite{dingle:2015} for RNA sequences of different
lengths. Lines are fits to data of the form $v_i=v_i^{\infty}-b_ie^{-c_iL}$,
from which the asymptotic values of the two versatilities $v_i^{\infty}$ are
extrapolated.}
\label{fig:Fig4}
\end{figure}

Consider long RNA sequences ---irrespective of their composition--- folded into
secondary structures. It has been shown that paired and unpaired sites admit on
average a different amount of mutations in a given NC, that is, they differ in
neutrality. Asymptotically, the overall neutrality of a phenotype can be well
described by two values, each corresponding to one of the structural
elements~\cite{huynen:1996b,reidys:2001}. In this vein, we consider now a
simplified model with two versatility values: one for paired ($v_p$) and one for
unpaired ($v_u$) sites (with $1\le v_p<v_u\le k$ for an alphabet of $k$ letters).
As neutrality, site versatility depends in principle on many factors other than
whether the corresponding base forms a bond. Nevertheless, we do observe that, on
average, versatilities associated to paired sites are significantly smaller than
those associated to unpaired ones. Interestingly, previous works have identified
a clear correlation between RNA secondary structure elements (stems and loops)
and nucleotide composition~\cite{schultes:1997,smit:2006}, giving indirect
support to our approximation.

The two-versatility model was introduced \cite{manrubia:2017} to argue for a
log-normal distribution of the abundance of RNA sequences in NNs. {More precisely,
the number} of RNA secondary structures with a given number $\ell$ of paired sites
{can be} shown to be (in the limit $L\to\infty$) proportional to a {Gaussian
function of $\ell$} with mean $\mu L-\mu_0$ and standard deviation $\sigma
L^{1/2}-\sigma_0L^{-1/2}+O(L^{-3/2})$ ($\mu=0.28647$, $\mu_0=1.36502$,
$\sigma=0.25510$, $\sigma_0=0.00713$). In virtue of \eqref{eq:estimatedsize}
{and the fact that, within the two-versatility model,
$S=v_p^{\ell}v_u^{L-\ell}$---hence $\ell\propto\log S$---this} immediately leads
to a log-normal distribution of $S$ with mean and standard deviation
\begin{align}
\mu_L &=L(\ln v_u-\mu)+\mu_0+O\left(L^{-1}\right),
\label{eq:meanL} \\
\sigma_L &=2\ln(v_u/v_p)\left(\sigma L^{1/2}-\sigma_0L^{-1/2}\right)
+O\left(L^{-3/2}\right).
\label{eq:sigmaL}
\end{align}

In order to test this two-versatility model we will use the data of
Ref.~\cite{dingle:2015} ---a collection of estimates of the abundance
distribution of RNA secondary structures obtained by sampling random sequences
of lengths in the range $L=20$--126. The resulting distributions are
proportional to $Sp(\ln S)$ but, if $p(\ln S)$ is a normal distribution with
mean $\mu_L$ and standard deviation $\sigma_L$, then so is $Sp(\ln S)$, with
the same standard deviation but a shifted mean {$\mu_L+\sigma_L^2$}. Fitting
Gaussian functions to these data yields $\mu_L$ and $\sigma_L$. Then,
through Eqs.~\eqref{eq:meanL}, \eqref{eq:sigmaL} we can infer the corresponding
versatilities $v_p$, $v_u$ ---which appear in Fig.~\ref{fig:Fig4}. This plot
suggests that these versatilities have well defined asymptotic values for
$L\to\infty$, namely $v_p=1.17\pm 0.08$, $v_u=2.79\pm 0.08$. For comparison,
the average versatilities obtained from our data for $L=16$ are
$v_p^{\text{av}}=1.11$, $v_u^{\text{av}}=2.37$.

A caveat is in order here. The results of \cite{dingle:2015} correspond to the
abundance of phenotypes, no matter how many NCs they have, whereas, strictly
speaking, the two-versatility model can only be applied to the latter. The
surprising agreement of the extrapolated versatilities with those directly
obtained from the data for $L=16$ suggests that for $L$ large, either phenotypes
are broken into few NCs, or one of these components is much larger than the
others and dominates the abundance of the phenotype. The existence of genetic
correlations in NCs seems to cause both effects~\cite{ahnert:2017}. Even for
short RNA and HP sequences, the largest connected component of a phenotype grows
linearly with the abundance of the phenotype, while the number of components either
diminishes with phenotype abundance~\cite{gruner:1996b} or remains mostly
independent~\cite{greenbury:2016}. Therefore, the largest NC becomes more
dominant the larger the phenotype, so that the latter is well approximated by a
single component. In consequence, the distribution of phenotype abundances is
asymptotically equivalent to the distribution of NCs abundances.

\begin{table*}
\caption{Data corresponding to the exhaustive {enumeration} of {phenotypes in}
multiple GP maps. The first column lists the maps studied and some of its
quantitative properties: total number of phenotypes, number of non-empty (NE)
phenotypes ({this quantity resulting from folds with non-negative energy and
the large number of degenerated genotypes that are discarded, see Appendix~A}),
number of sequences assigned to a unique phenotype (UaS), average abundance of
phenotypes $S_{av}$, total number of neutral components (NCs), and fraction of
non-functional sequences ($f_{\emptyset}$). Non-compact HP20 (n-c HP20) is
included to compare with n-c HP20 with minimal contact maps (n-c HP20 ${\cal
S}$) as phenotypes (a distribution of phenotype abundances for n-c HP20 can be
found in~\cite{shahrezaei:1999}). $^1$Data obtained with two energy parameters,
$U(H,H)=-2.3$ and $U(H,P)=U(P,H)=-1$. $^2$Data from~\cite{holzgrafe:2011}.}
\label{tab:exhaustive}
\begin{center}
  \begin{tabular}{|l|r|r|r|r|r|r|}
    \hline
    \multicolumn{1}{|c|}{\bf Model} &
    \multicolumn{1}{|c|}{Phenotypes} &
    \multicolumn{1}{|c|}{NE phenotypes} &
    \multicolumn{1}{|c|}{UaS} &
    \multicolumn{1}{|c|}{$S_{av}$} &
    \multicolumn{1}{|c|}{NCs} &
    \multicolumn{1}{|c|}{$f_{\emptyset}$} \\
    \hline
    RNA30 GC & 240,944,076 & 432,221 & 1,073,725,603 & 2,484.2 & 68,389,814 & 0.0000151 \\
    RNA16 ACGU & 5,223 & 648 & 1,712,323,320 & 2,642,474 & 23,092 & 0.601 \\
    compact HP30 & 13,498 & 13,498 & 187,212,435 & 13,869.6 & 362,221 & 0.826 \\
    compact HP30$^1$ & 13,498 & 13,498 & 258,434,457 & 19,146.1 & 1,986,907 & 0.759 \\
    n-c HP30$^2$ & 784,924,528,667 & 2,333,498 & 22,466,621 & 9.63 & 3,732,449 & 0.979 \\
    n-c HP20 & 41,889,578 & 5,310 & 24,900 & 4.69 & 6,586 & 0.976 \\
    n-c HP20 ${\cal S}$ & 910,971 & 54,818 & 292,732 & 5.34 & 62,379 & 0.721 \\
    \toyLIFE & $2^{214} \simeq 2.63 \times 10^{64}$ & 775 & 134,400,450 & 173,419.9 & 1,523,544 & 0.9999 \\
    \hline
  \end{tabular}
\end{center}
\end{table*}

The improvement of the fit upon increasing length can be indirectly inferred
from the data of Ref.~\cite{dingle:2015}.
The fits of Gaussian functions to these data are more
accurate than the one of Fig.~\ref{fig:Fig3}(a) (see Appendix~D and Fig.~\ref{fig:FigS2}),
and show that the log-normal behaviour of $p(S)$ is what should be expected for
long sequences.

We can apply the two-versatility model to our results with GC-RNA. The effective
versatilities are $v_p=0.75$ and $v_u=1.32$ (from the data we obtain the exact
value $v_p=1$ and the average $v_u^{\text{av}}=1.43$). As in the case of four-letter
RNA (c.f.~Fig.~\ref{fig:Fig4}), the values of $v_p$ for short lengths are
unphysical ($v_p<1$). This notwithstanding, effective versatilities are not too
far from the average ones, providing an indirect support to the fact that the
log-normal distribution for this model has a mean close to 1 ---explaining why
only the right branch is observed.

\section{Phenotype definition, alphabet size, and navigability of genotype
  spaces}

\begin{figure}
\includegraphics[width=88mm,clip]{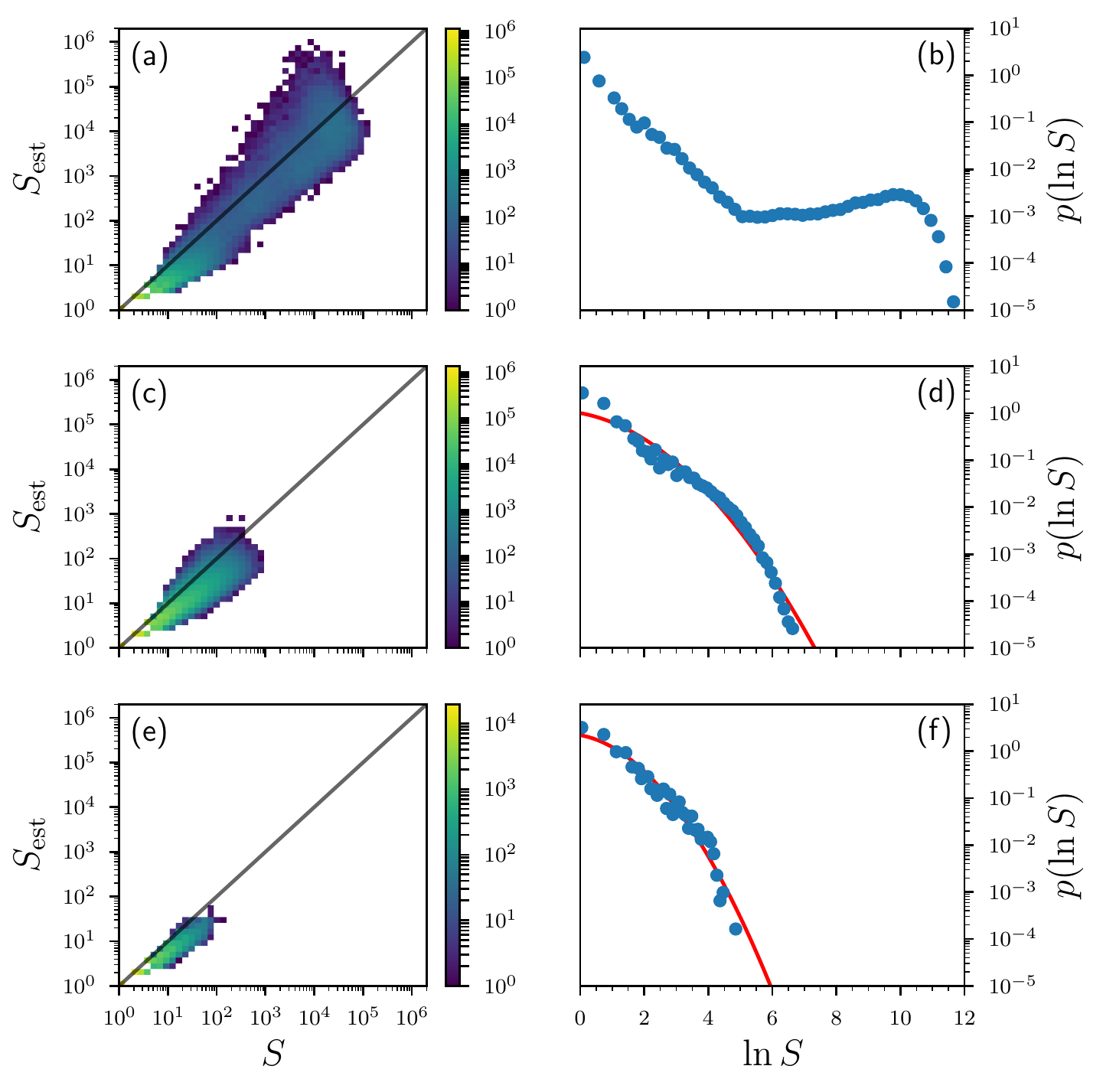}
\caption{(a, c, e)- Log-log-log histograms of the estimated abundance
  $S_{\text{est}}$ {\it versus} actual abundance $S$ of the NCs of different HP
  versions. (b, d, f) NCs abundance distributions. (a, b) Compact HP 5$\times$6
  with $U(H,H)=-2.3$ and $U(H,P)=U(P,H)=-1$, (c, d) non-compact HP30 with
  $U(H,H)=-1$, and (e, f) non-compact HP20 $\cal S$ (based on minimal contact
  maps) with $U(H,H)=-1$.}
\label{fig:Fig5}
\end{figure}

Figure~\ref{fig:Fig2} suggests that the goodness of the phenotype abundance
estimation \eqref{eq:estimatedsize} might depend on the specific GP map. While it
works amazingly well for four letter RNA, it is not that good for compact HP or
\toyLIFE, which have similarly large NCs. Indeed, high accuracy in that prediction
implicitly relies (i) on the existence of a clear-cut quantitative relationship
between sequence sites and structural elements ---which is mediated by a
consistent definition of phenotype, and (ii) on the presence of a giant NC in
phenotypes. The latter seems essential for the abundance of phenotypes to follow
a {\it bona fide} log-normal distribution. Though the relationship between
sequence and structure is unequivocal for RNA, it depends on the definition of
phenotype in various versions of the HP model (see Appendix~A), becomes
unavoidably cryptic for
\toyLIFE, and might be hard to define in GP maps lacking an easy representation
of genotypes as sequences~\cite{ibanez:2014}. On the other hand, a comparison of
the distribution of abundances in two- and four-letter RNA indicates that the
larger the alphabet the larger the components of phenotypes and the better
defined the log-normal distributions. These observations are in full agreement
with results for the HP model, as shown in the following. 

Figure~\ref{fig:Fig5} illustrates the performance of versatility and abundance
distributions for three additional definitions of phenotype in HP models: compact
HP30 with two parameters for energy (Fig.~\ref{fig:Fig5} (a) and (b)),
non-compact HP30 ((c) and (d)) and non-compact HP20 with phenotypes defined
through {\it minimal} contact maps ((e) and (f)) that is, the set
${\cal S}_{ij}$ formed by those pairs with a nonzero contribution to the folding
energy, ${\cal S}_{ij}=\{i, j \, | \, U(\sigma_i,\sigma_j) C_{ij} <0\}$.

Initially, the HP model was implemented in its compact version for computational
tractability: notice that the number of different two-dimensional
{conformations} in compact HP30 is $10^8$-fold smaller than in
non-compact HP30 (Table~\ref{tab:exhaustive}). Compact HP versions actually impose
unrealistic spatial constraints: two residues $i$ and $j$ can be forced to be in
contact without having an associated interaction energy, that is $C_{ij}=1$, but
$U(\sigma_i,\sigma_j)=0$. Spatial restrictions may therefore assign to a unique
phenotype (or NC thereof) sequences whose affiliation easily changes under more
natural phenotype definitions~\cite{irback:2002}. This has an immediate effect on
abundance distributions, as Fig.~\ref{fig:Fig5}(b) shows: besides a decrease at
small NC sizes, the distribution develops a bump at high abundances. The
non-compact versions of HP are difficult to explore exhaustively due to the
astronomically large number of possible phenotypes~\cite{holzgrafe:2011}. Still,
phenotypes are free from spatial constraints and, as a result, abundance
distributions can be fit with a log-normal function (Fig.~\ref{fig:Fig5}(d),
(f)). These distributions are qualitatively similar to that obtained for GC-RNA,
though NCs are significantly larger in the latter. Smaller NCs could be expected
if, instead of the Vienna Package to fold RNA sequences, a model with few energy
parameters (such as, e.g., Nussinov algorithm for loop
matching~\cite{nussinov:1978}) is used. 

In either compact or non-compact realisations, folding is calculated by using
one~\cite{holzgrafe:2011} or two~\cite{li:1996} nonzero energy parameters,
examples being $U(H,H)=-1$, as in Fig.~\ref{fig:Fig2}(c)) or $U(H,H)=-2.3$, and
$U(H,P)=-1$, e.g., as in Fig.~\ref{fig:Fig5}(a)). Genotypes in these HP models
can typically be mapped to more than one phenotype. Traditionally, these
degenerated genotypes are discarded, since they have been interpreted as the
analogues of intrinsically disordered proteins, and therefore devoid of function.
This convention results in one of the most concerning features of classical HP
models~\cite{buchler:1999}, where an astonishingly large fraction of sequences
are systematically not assigned to phenotypes, yielding {empty
  phenotypes and many}
small and highly
fragmented {ones} (see Table~\ref{tab:exhaustive} for representative
examples). It is important to remark that a high fraction of non-functional
sequences does not necessarily imply that phenotypes are small and isolated,
since other models ---where the small fraction of functional sequences is not due
to degeneration--- do have large and easily navigable
phenotypes~\cite{ciliberti:2007,matias-rodrigues:2009,arias:2014}.

Adding more energy parameters serves to disambiguate the assignation of genotypes
to phenotypes, though the increase in the fraction of sequences assigned to
phenotypes is however minor (compare the two compact HP30 versions in
Table~\ref{tab:exhaustive}). Phenotypes defined through contact maps are closer
analogues of RNA secondary structure (as in our example with non-compact HP20):
contact maps appear as a more natural definition of phenotype that furthermore
reduces about 40-fold the number of different phenotypes and notably decreases
sequence degeneration (Table~\ref{tab:exhaustive}). Also, degeneration diminishes
significantly when the size of the alphabet grows. In a systematic study with
sequences of length $L=25$, degeneration is halved when going from two- to
four-letter alphabets, and it reaches a few percent for 20-letter
representations~\cite{buchler:2000}. Concomitantly, phenotypes become larger and
more connected.  

The fact that most phenotypes are small, weakly connected and even difficult to
navigate in classical HP models~\cite{holzgrafe:2011} raises doubts on their
relevance for evolutionary dynamics, speaking in favour of more complex but also
more realistic scenarios~\cite{buchler:1999}, and certainly supporting
non-compact versions of lattice protein models~\cite{irback:2002}. In agreement
with the above, the definition of phenotype critically affects the distribution
of abundances, which changes from decreasing functions for two-letter
alphabets~(as in Fig.~\ref{fig:Fig5}) to functions with a maximum and a fat tail
for 20-letter, compact versions~\cite{buchler:1999,li:2002}. Independent studies
suggest that minimal alphabets are not optimal in an evolutionary
sense~\cite{gardner:2003}, further supporting the limited applicability of
two-letter models, especially to draw conclusions on evolutionary dynamics.
Unfortunately, an exhaustive study of non-compact lattice protein models with
more than two letters is, as of today, computationally unfeasible. 

\section{Conclusions}

The vastness of genotype spaces prevents a complete characterisation based on
computational approaches. A look at Table~\ref{tab:exhaustive} suffices to
illustrate the astronomically large numbers involved in calculations with
sequences of length well below that typically found in biochemical processes. The
data generated to analyse the different models in this contribution reaches 0.5TB
and, as their diversity shows, would be of limited use in the absence of an
accompanying theory. Therefore, an understanding of the structure of realistic GP
maps demands further theoretical developments that can be extrapolated to
arbitrarily long sequences. We have shown that the definition of useful
quantities such as versatility allows for reliable estimations of the abundance of
phenotypes and for the derivation of the expected distribution. Knowledge of the
asymptotic values $v_p$ and $v_u$ yields that distribution in RNA of any length,
as well as an estimation of the number of genotypes folding into an arbitrary
(typical) structure. Similar derivations should be possible for other GP maps
endowed with consistent definitions of phenotype.

\acknowledgments

K Dingle, E Ferrada, Ch Holzgr\"afe, A Irb\"ack and A Louis are gratefully
thanked for sharing their data with us. We acknowledge financial support by the
Spanish Ministerio de Econom\'{\i}a y Competitividad and FEDER funds of the EU
through grants VARIANCE (FIS2015-64349-P; PC, JAC) and ViralESS (FIS2014-57686-P;
JAGM, SM).

\bibliographystyle{eplbib} 
\bibliography{bibliography}

\section{APPENDIX A: Phenotype definitions}

There is no unique, unambiguous or optimal definition of phenotype. In general,
it is an environment-dependent quantity that may take a variety of forms in
computational models as those studied in this work. On the positive side, most
statistical properties of the genotype-phenotype map are independent (or weakly
dependent) on the precise definition.


\subsection{RNA}
 
The most used energy model for RNA secondary structure
folding~\cite{turner:2010} is based on experimentally measured energy
contributions of hydrogen bonds and stacking interactions between paired and
adjacent nucleotides, energetically unfavourable unpaired regions, such as
hairpin loops, bulges, internal loops and multiloops, and specific favourable
contributions, such as GNRA tetraloops. The minimum free energy structure of a
given RNA chain (genotype) can be thus determined by applying Zuker's dynamic
programming algorithm \cite{zuker:1981} with the given energy model.

RNA GP maps are constructed by assigning to each genotype its minimum free
energy secondary structure as phenotype. In cases when the folding energy is
negative, there might be still more than one secondary structure with the same
(minimum) energy. In general, when a genotype can be assigned to more than one
phenotype it is said to be {\em degenerated}. RNA folding algorithms, however,
randomly select one of those minimum energy structures and assign the genotype
to that phenotype. When the folding energy is non-negative, the corresponding
phenotype is then the empty phenotype (the open structure with no base pairs)
and the genotype is considered as {\em non-functional}. 

\subsection{HP model}

HP models consider chains of hydrophobic and polar amino acids (genotypes)
placed in a two or three dimensional lattice, which imposes spatial folding
constraints. Depending on the size of that lattice the model can be {\em
compact} or {\em non compact}. In compact models the lattice size is finite,
and in the case considered in this work it equals the protein length. The size
of the lattice forces proteins to fit into a given space applying the rationale
that proteins preferentially fold into globular structures. This assumption
drastically reduces the number of possible phenotypes. On the other hand, {\em
non-compact} models consider an infinite lattice, where possible conformations
are only limited by the protein backbone and the spaces occupied by other amino
acids.

An often used definition of phenotype in HP models is the conformation or {\em
path} in which it has the lowest free energy. Phenotype is therefore
characterized by the path followed by the backbone of the sequence on the
lattice (after symmetries are eliminated), regardless the number or position of
actual contacts between residues. This is the basic definition we use for HP
compact and non-compact models in the main text if no variant is specified. In
this context, it is usually assumed (at odds with RNA) that if a protein can
fold into more than one conformation with the same minimum energy the phenotype
is ambiguous and the genotype is {\em degenerated}. Classical approaches
consider degenerated genotypes as {\em non-functional}, and therefore discard
the corresponding genotypes, which are not included in further analyses. Given
the small number of parameters involved in energy calculations, degeneration
occurs very frequently, leading to a perhaps surprisingly high number of
non-functional sequences in HP models (a quantity that includes folds with
non-negative energy but also folds with negative energy if they are
degenerated).  
As a result, most of the possible phenotypes are empty, just because compatible
genotypes happen to yield more than one minimum free energy conformation.
Non-compact HP models are especially affected by degenerated genotypes due to
the conformational freedom intrinsic to non-interacting parts of the genotype.

Degeneration is a somehow artefactual property that can be reduced under more
realistic definitions of phenotype, for example by using a different energy
model. The simplest model (HH) considers only interactions between hydrophobic
amino acids, assigning an energy of -1 to each H-H interaction. An alternative
energy model used in this work also assigns an energy to hydrophobic-polar
interactions ($H-H = -2.3$ and $H-P = -1$). Another possibility is to use {\em
minimal contact maps} to define phenotypes. The minimal contact map of a
sequence minimizes the free energy and also the number of non energetically
favourable contacts (energy $\leq 0 $). This definition can account for the
contribution of non favourable interactions (which can be weak compared with
hydrophobic interactions, but should not be ignored) and discards
conformational changes due to non-interacting parts of the genotype.
Note that this definition is similar to RNA secondary structure models defined
by sets of paired and unpaired regions.


\section{Appendix B: Computing versatilities using connected components instead of
  the whole neutral network}

In the main text, we consider the different connected components of a
phenotype to be different phenotypes in order to estimate their abundances. The
reason for this is that if we compute the versatilities lumping all connected
components together, the abundance of the phenotype is typically greatly
overestimated, unless one of the connected components dominates the neutral
network---in which case lumping or not makes little difference.

To understand why this is so we can use a very simple example. Suppose a
phenotype that is codified by the following sequences:
\begin{center}
\begin{tabular}{cccc}
  000000 & 010000 & 100000 & 110000 \\
  001111 & 011111 & 101111 & 111111
\end{tabular}
\end{center}
Clearly, each row belongs to a different connected component, because all
genotypes in the upper row end in 0000, and all the genotypes in the lower row
end in 1111 (for a GP map such as toyLIFE, this situation is not uncommon).
Calculating site versatilities with our formula (c.f.~Eq.~(2) in main text) for
each separate connected component we obtain
\begin{equation}
  v_1=v_2=2, \quad v_3=v_4=v_5=v_6=1.
\end{equation}
Estimating the size of these components with Eq.~(1) (main text) then leads to
the correct result $S=4$ for both of them. However, if we lump together all
genotypes into a single phenotype Eq.~(2) (main text) estimates site
versatilities as $v_i=2$ \emph{for all sites,} and so the size estimate
obtained from Eq.~(1) will be $2^6=64$, $8$ times larger than the actual
abundance.

From this example one can also understand why, if one of the connected
components is much larger than the rest, the distortion of lumping all
components together will have a small effect on our estimate of phenotype
abundance.

\section{APPENDIX C: Estimating phenotype abundance using compositional entropy}

In the main text, we present one way to estimate the abundance of a
phenotype by defining the versatility of a
site $i$ as
\begin{equation}
v_i=\frac{1}{M_i}\sum_{\alpha=1}^km_{\alpha,i}, \quad
M_i\equiv\max\{m_{1,i},\dots,m_{k,i}\},
\end{equation}
where $m_{\alpha,i}$ is the number of genotypes of the phenotype in
which the letter $\alpha$ appears at site $i$, and $k$ is the size of
the alphabet. The estimated abundance of a phenotype $S_{\text{est}}$ would then be given by
\begin{equation}
  S_{\text{est}}=\prod_{i=1}^L v_i.
\end{equation}

Previous work \cite{li:1996,larson:2002} had proposed a different
formula for the estimation of phenotype abundance using compositional
entropy: 
\begin{equation}
  p_i=- \sum_{\alpha=1}^k \frac{m_{\alpha,i}}{M_S}
  \ln \left(\frac{m_{\alpha,i}}{M_S} \right), \quad
M_S\equiv\sum_{\alpha=1}^k m_{\alpha,i},
  \label{eq:entropy}
\end{equation}
The abundance of the
phenotype would then be computed as
\begin{equation}
  S_{\text{est}}= \prod_{i=1}^L \exp(p_i).
  \label{eq:estimatedsize2}
\end{equation}

\renewcommand{\thefigure}{S\arabic{figure}}
\setcounter{figure}{0}

\begin{figure}
\includegraphics[width=88mm,clip]{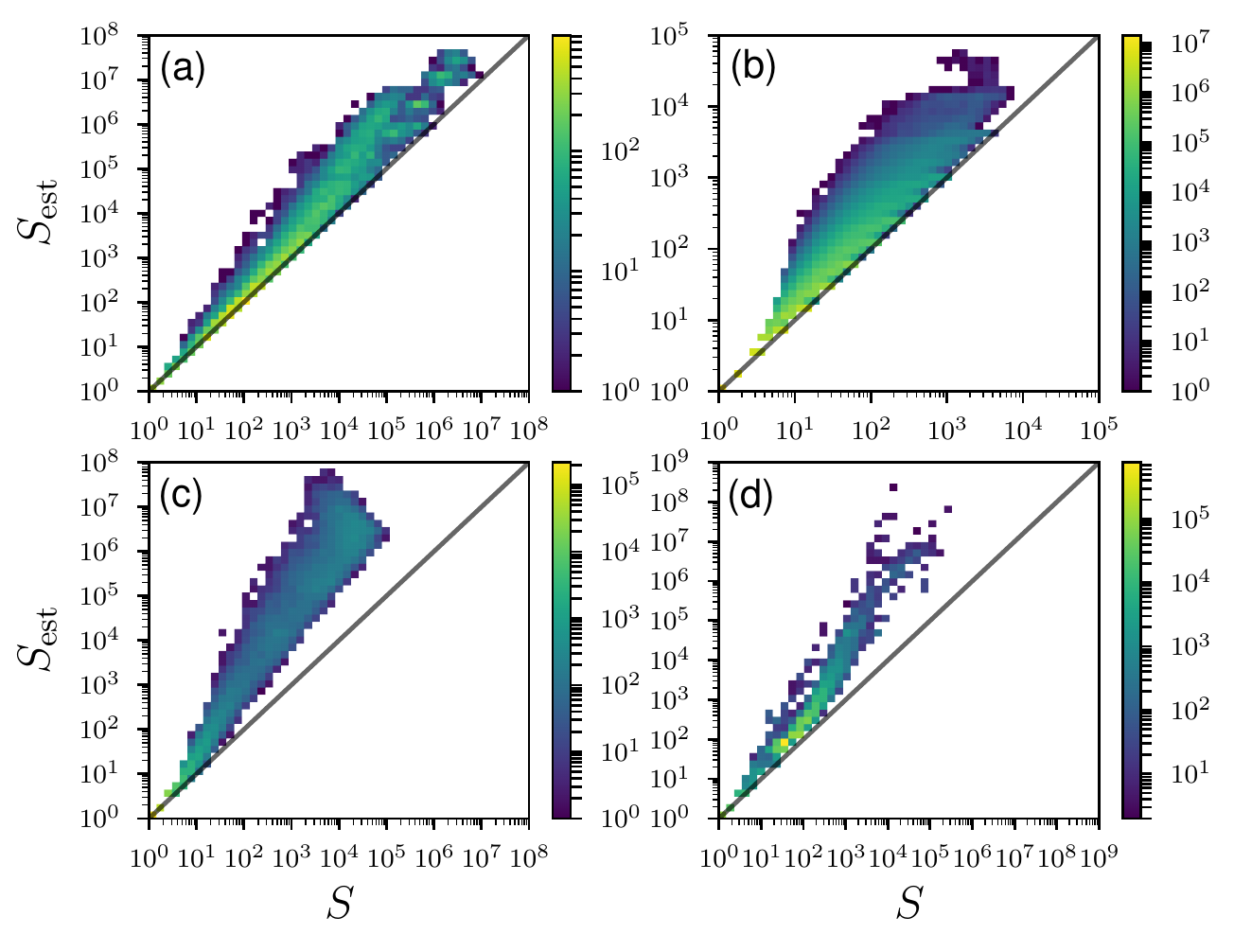}
  \caption{Log-log-log histograms of the estimated abundance ($S_{\text{est}}$)
  calculated using compositional entropy (see main text), {\it versus} actual abundance ($S$)
  of the connected components of different GP maps: (a) four-letter RNA of length
  $L=16$, (b) two-letter GC-RNA of length $L=30$, (c) compact HP model
  5$\times$6 with $U(HH)=-1$, and (d) \toyLIFE{} for two genes.}
\label{fig:FigS1}
\end{figure}

In Figure \ref{fig:FigS1} we test how well this formula predicts
phenotype abundance, comparing actual abundances with estimates for
the four models studied in the main text, namely (a) RNA sequences of
length $L=16$, (b) two-letter GC-RNA sequences of length $L=30$, (c)
compact HP proteins folding on a 5$\times$6 lattice ($L=30$), and (d) \toyLIFE{} genotypes
with two genes ($L=40$).

In all cases, compositional entropy overestimates phenotype
abundance. Worse, the prediction seems to follow a power law
$S_{\text{est}} \propto S^{\gamma}$, with $\gamma >1$ and depending on
the model. This means that the most abundant phenotypes are
overestimated the most. This is an undesirable property, as abundant
phenotypes are the ones that do appear in nature \cite{dingle:2015}.

We can try to correct this method by taking into account correlation
between sites, estimating the logarithm of phenotype abundance by
\begin{equation}
\ln
S_{\text{est}}=-\frac{1}{L-1}\sum_{\alpha,\beta=1}^k\sum_{i=1}^L\sum_{j>i}p_{ij}^{\alpha\beta}
\ln p_{ij}^{\alpha\beta},
\label{eq:estimatesize3}
\end{equation}
where $p_{ij}^{\alpha\beta}$ is the probability that the symbol
$\alpha$ appears at site $i$ and symbol $\beta$ appears at site
$j$. If $p_{ij}^{\alpha\beta}=p_i^\alpha p_j^\beta$, that is, if there
are no correlations, this formula
reduces to \eqref{eq:estimatedsize2}. The results (not
shown) are very similar to those in Fig.~\ref{fig:FigS1}, implying
that correlations are not the source of the mismatch between predicted
and actual abundances. In summary, compositional entropy is not a good
method to estimate phenotype abundance.

\begin{figure}
\includegraphics[width=88mm,clip]{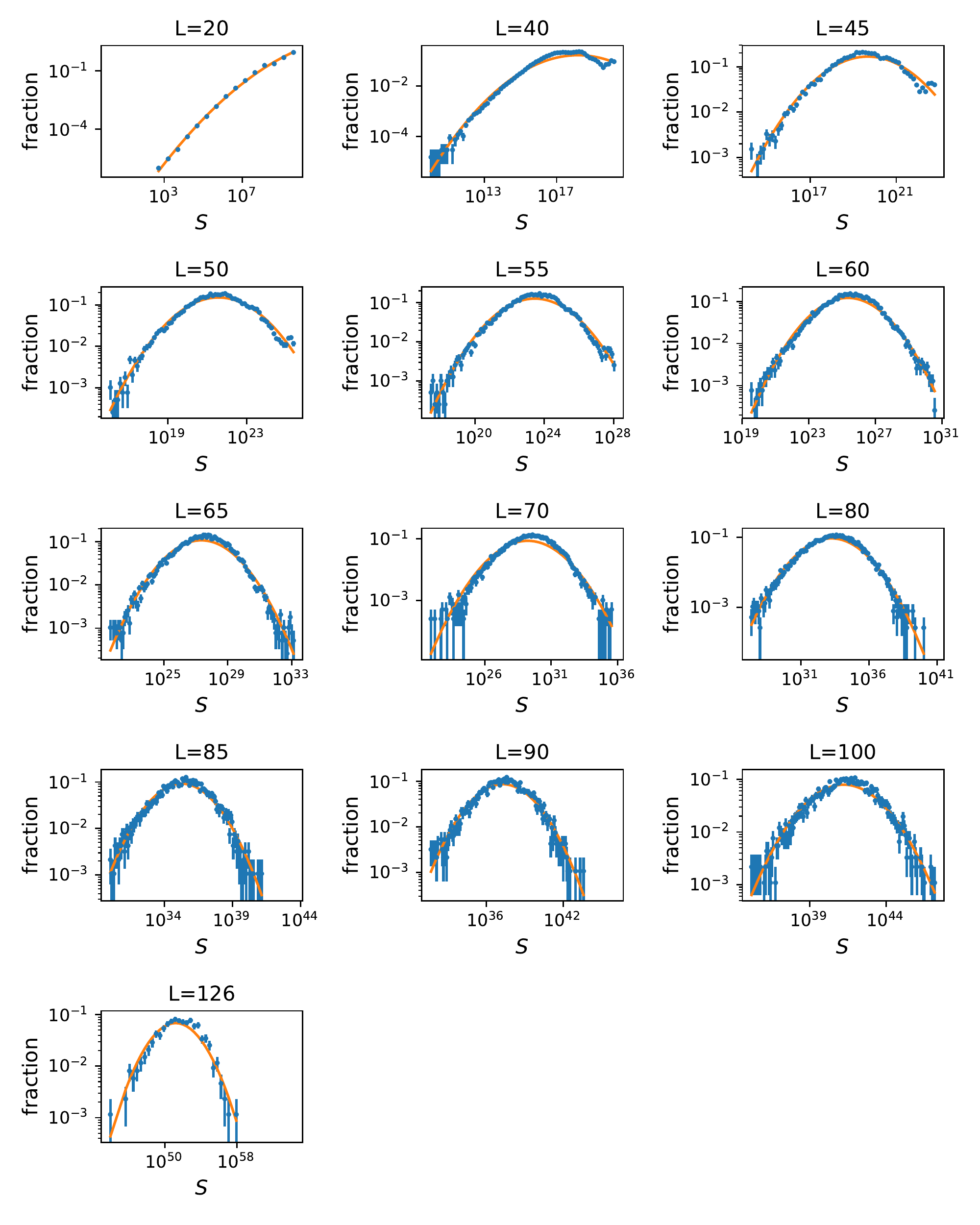}
  \caption{Log-normal fits to the abundance distributions of RNA
    secondary structures from Ref.~\cite{dingle:2015},
    for RNA sequences of lengths from $L=20$ to $L=126$. The $x$-axes
    represent phenotype abundance $S$ (in logarithmic scale), while the
  $y$-axes represent the fraction of phenotypes of abundance $S$ when
    sampling at random among all possible genotypes. These fractions
    are proportional to $Sp(\ln S)$.}
\label{fig:FigS2}
\end{figure}

\section{APPENDIX D: Fitting log-normal functions to abundance distributions of
RNA secondary structures}

Ref.~\cite{dingle:2015} presented a collection of estimates of the
abundances of RNA secondary structures for sequences of length $L=20$
to $L=126$, obtained by random sampling of RNA sequences. The
resulting distributions (shown in Fig.~\ref{fig:FigS2} with Gaussian
fits) are proportional to $Sp(\ln S)$, since the
process of choosing a phenotype of abundance $S$ (with probability
proportional to $S$) is weighted by the number of phenotypes with that
abundance, which is given by $p(\ln S)$. This
distribution will follow a normal distribution if $p(\ln S)$ does as
well, but with a shifted mean (Fig.~\ref{fig:FigS2}). This shift
explains why the distribution of abundances of secondary structures
for $L=20$ is increasing, contrasting with the rest of distributions
we show in the main text. 


\end{document}